\documentclass[aps,prb,groupedaddress,amsmath,twocolumn,showpacs,floatfix]{revtex4}
\DeclareMathOperator{\sech}{sech}
\DeclareMathOperator{\erfc}{erfc}

\usepackage{graphicx}
\begin{document}

%\preprint{}


\title{Finite-size fluctuations and photon statistics near the
  polariton condensation transition in a single-mode microcavity}


\author{P. R. Eastham and P. B. Littlewood}
%\thanks{}
\affiliation{Theory of Condensed Matter Group, Cavendish Laboratory,
  Madingley Road, Cambridge, CB3 0HE. U.K.}


\date{\today}

\begin{abstract}
  We consider polariton condensation in a generalized Dicke model,
  describing a single-mode cavity containing quantum dots, and extend
  our previous mean-field theory to allow for finite-size
  fluctuations. Within the fluctuation-dominated regime the
  correlation functions differ from their (trivial) mean-field
  values. We argue that the low-energy physics of the model, which
  determines the photon statistics in this fluctuation-dominated
  crossover regime, is that of the (quantum) anharmonic
  oscillator. The photon statistics at the crossover are different in
  the high- and low- temperature limits. When the temperature is high
  enough for quantum effects to be neglected we recover behavior
  similar to that of a conventional laser. At low enough temperatures,
  however, we find qualitatively different behavior due to quantum
  effects.
\end{abstract}

%\keywords{}

\maketitle


\section{Introduction}
\label{sec:introsec}

Microcavity
polaritons\cite{hopfield58,weisbuch92,skolnick98,savonapol98,kavokinbook}
are quasiparticles which form in wavelength-scale optical cavities
containing dielectrics. They are mixed modes formed from cavity
photons and dielectric excitations such as excitons. Since they are
part photon polaritons are bosons, and are thus candidates for Bose
condensation.

Experimental results on pumped microcavities continue to be linked to
polariton condensation\cite{deng03,richard05,dang98,senellart99}. The
basic result is a threshold behavior of the luminescence intensity
from a driven microcavity, while other features seen include
non-thermal correlation functions for this luminescence, along with
spatially and spectrally localized emission.

The central characteristic of Bose condensation is the generation of
many-particle coherences which, in polariton condensation, appear in
the electromagnetic field. The existence of polariton condensation
remains controversial because polariton condensation is not the
only phenomenon we can associate with coherent photons in
microcavities. Most straightforwardly, if the cavity is driven into
the weak-coupling regime one expects conventional lasing, and this is
thought to be the correct interpretation of early claims for polariton
condensation\cite{pau96,fan97}. While the more recent experiments
cannot be straightforwardly attributed to conventional lasing, more
exotic alternatives to polariton condensation, such as polaritonic
lasing\ \cite{laussy04}, remain.

While there may appear to be many different routes to optical
coherence in microcavities the relationships amongst these routes are
not clear. Polariton condensation, polariton lasing, and conventional
lasing are often assumed to be fundamentally distinct, but could
equally well be related phenomena in different parameter regimes. This
view is supported by recent work showing that adding decoherence
processes to a mean-field theory of polariton
condensation\cite{eastham00,eastham01,easthamthesis} leads to a
crossover from condensation to conventional
lasing\cite{szymanska02,szymanska03}. It is also suggested by the
fundamental connections between equilibrium and non-equilibrium phase
transitions.  The best known of these connections is between laser
theory and the Landau theory of second-order phase
transitions\cite{haken75,risken70}, but we note also recent work
connecting the critical behavior of parametric oscillators and
ferromagnets\cite{drummond05}, and a treatment of the ideal Bose gas
along the lines of laser theory\cite{scully99}.

In this paper we investigate fluctuations close to polariton
condensation, and how they affect the photon statistics. In general
there are different regimes for the dominance of fluctuations: in a
very large system at low density of excitation the thermal
equilibrium transition is of the BEC variety,\cite{keeling04,keeling05}
so that spatial fluctuations are important.  But since the polariton
mass is very small (because of the large ratio between the wavelength
of light and the typical exciton radius and exciton separation), at
modest densities a BCS-like mean-field regime occurs and spatial
fluctuations are small.  Here the dominant fluctuations may be due to
the finite size -- or generically the finite population -- of the
system. This is very often also the case for conventional
lasers\cite{haken75} (for similar reasons). 

Here we consider finite-size fluctuations in isolation by studying a
single-mode model microcavity. We obtain an approximate form for the
free energy of the model by neglecting quantum effects. This form can
be interpreted as the classical probability distribution for the
intensity of the cavity field, and used to obtain all the static
correlation functions of the cavity photons. It is identical to the
accepted form for the intensity distribution near the onset of lasing,
so we argue that within our approximations polariton condensation and
lasing are not distinguished by the qualitative behavior of the static
correlation functions. However the parameters in the intensity
distribution are associated with different physics in the two
theories, so there remains room for quantitative distinctions between
them.

While neglecting quantum effects in a theory of condensation leads to
the same intensity distribution predicted by classical laser theory,
this approximation fails for a condensate at low temperatures. When
the temperature becomes comparable with or less than the interaction
energy of two photons the finite level spacing affects the
correlation functions.  We shall see that this leads to behavior for
the correlation functions of a condensate at low temperature which is
qualitatively different from the predictions of standard laser theory.
This difference is not surprising because standard laser theory is a
classical approximation, controlled by the photon number at threshold.

Our analysis concerns the thermal equilibrium of a simplified model of
a microcavity. Although in some parameter regimes current experiments
are far from thermal equilibrium, there are several reasons to study
the equilibrium behavior. The above-threshold luminescence in some
recent experiments\cite{deng03} is suggestive of thermal equilibrium,
and as microcavities continue to develop\cite{tawara04,pawlis04}
experiments can be expected to reach states closer to thermal
equilibrium.  Furthermore, the behavior close to equilibrium is
expected to be similar to that in equilibrium, and an understanding of
the equilibrium physics provides the basis for developing
non-equilibrium theories.  Finally, the qualitative behavior we
exhibit here, in particular the functional forms of the correlation
functions, derives from the structure of the effective theory
describing the collective variables.  This structure may be
independent of whether the theory describes an equilibrium system such
as a conventional condensate or a non-equilibrium system such as a
conventional laser.

The remainder of this paper is organized as follows. We begin with an
outline of the main results in section\ \ref{sec:outline}, which
compares the classical laser, the classical and quantum critical
fluctuations of the polariton condensate in terms of the anharmonic
oscillator. The rest of the paper provides a derivation of these
results, and a deeper quantitative analysis of the model.

In section\ \ref{sec:model} we present the model we consider for both
lasing and condensation, and give some necessary background on the
mean-field theory of polariton condensation.  Section\ 
\ref{sec:freeenergy} contains the general analysis resulting in the
free energy in the transition region, including a discussion of the
regime of applicability of the classical approximation. In section\ 
\ref{sec:corrfuns} we use these general results to analyze the phase
diagram and photon statistics of the condensate.  In section\ 
\ref{sec:laser} we briefly review the standard calculations of the
intensity distribution in a laser near threshold, and compare with our
results for the polariton condensate. In doing so, we note that
conventional laser theory and its approximation to a phase transition
relies on a ``large-N'' justification that is not usually exposed.
Section\ \ref{sec:numbers} contains numerical estimates of the size of
the fluctuation-dominated and quantum regimes in current condensation
experiments. In section\ \ref{sec:discussion} we discuss prospects for
systems with large quantum regimes, and the role of spatial
fluctuations. Finally, section\ \ref{sec:conclusions} summarizes our
conclusions.

\section{Outline of main results: the anharmonic oscillator}
\label{sec:outline}

We shall find that the classical laser and quantum/classical
condensate lie in the same universality class as the anharmonic
oscillator

\begin{equation} H= \alpha \phi^\dagger \phi +\gamma \phi^\dagger
\phi^\dagger \phi \phi. \label{eq:anharmosc}\end{equation} 
We choose $\phi$ to be normalized to obey the canonical Bose commutation
relation $[\phi,\phi^{\dagger}] = 1$.

The eigenstates of (\ref{eq:anharmosc}) are just the number states,
$\lvert n\rangle$, with energies $E(n)=(\alpha-\gamma) n+\gamma n^2 \approx
\alpha n+\gamma n^2$. Thus the partition function is
\begin{equation}Z=\sum_{n=0}^\infty e^{-\beta (\alpha n + \gamma
    n^2)},\label{eq:anharmz}\end{equation}
with $\beta = 1/k_BT$.

The parameter $\alpha$ is the tuning parameter through the transition.
At the mean-field level, minimization of the exponent in (\ref{eq:anharmz}) leads to
\begin{eqnarray}
n_{\mathrm{min}} &= 0 \;\;\; &{\rm for } \;\; \alpha > 0 \nonumber \\ \\
n_{\mathrm{min}} &= \frac{|\alpha|}{2 \gamma} \;\;\; & {\rm for } \;\; \alpha <
0. \nonumber
\end{eqnarray}
Expanding in terms of fluctuations, $\delta n = n- n_{\mathrm{min}}$, one
obtains (on the ``condensed'' side $\alpha < 0$) quadratic number
fluctuations controlled by $\gamma (\delta n)^2$. We shall choose the
parameter $\gamma \propto 1/N$, with $N$ growing with the system size;
the limit $N \to \infty$ gives the mean-field result.  We show below
that the Dicke model indeed gives rise to a partition function of the
form of (\ref{eq:anharmz}) after truncating higher order terms in the
exponent that are systematically smaller in powers of $1/N$ (where in
this case $N$ is the number of quantum dots in the cavity).

The summation in the partition function
is of course still over discrete quantum states $\lvert n\rangle$, but if the temperature is high
enough the discreteness is irrelevant and the sum can be replaced by an integral. We 
now use fields $\psi = \phi / \sqrt N$ rescaled by system size so that $\psi$ appropriately describes the classical electromagnetic
field intensity. Then one may write the partition function as
\begin{equation}
\label{eq:psi4}
Z \approx \int d\psi d \psi^* e^{-\beta N [ \alpha |\psi|^2 + \gamma N |\psi|^4]}
\end{equation}
(remembering that $\gamma N = O(1)$). We derive this form explicitly
for the Dicke model in section\ \ref{sec:freeenergy}.  We now remark
that the action in (\ref{eq:psi4}) is consistent with the steady-state
distribution from a Fokker-Planck equation for diffusion in a quartic
potential (see section\ \ref{sec:laser}), which is the conventional
approach to laser theory.  Of course the ``temperature'' there is a
fiction that is generated by couplings to (Markov) baths representing
the outside world. Furthermore there is no quantum limit for that type
of laser theory. In the classical limit the steady state laser and
the polariton condensate have the same scaling form.

The anharmonic oscillator thus encapsulates the basic results of the
two different models when we focus on the low energy physics. One
signature of the transition is the intensity-intensity correlation
function $g^{(2)}(0)$, which is straightforward to calculate for the
anhmarmonic oscillator:

\begin{equation} g^{(2)}(0)=\frac{\sum_n
n(n-1)e^{-\beta(\alpha n + \gamma n(n-1))}}{(\sum_n ne^{-\beta(\alpha
n + \gamma n(n-1))})^2} . \end{equation}

\begin{figure}[ht]
\begin{center}
\includegraphics[width=3in]{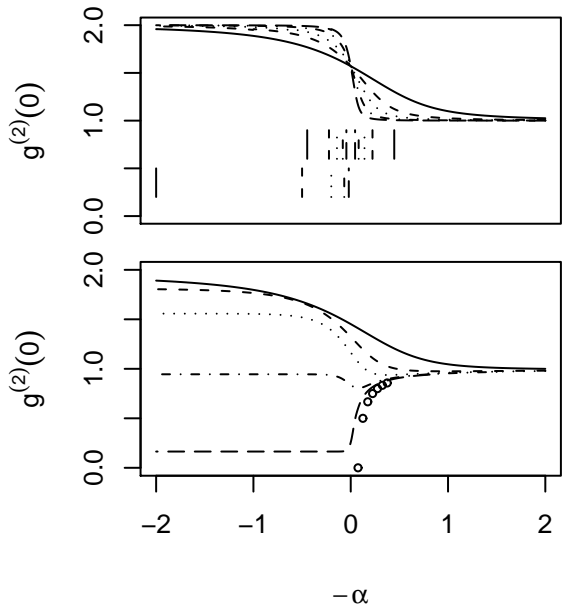}\end{center}
\caption{$g^{(2)}(0)$ as a function of $\alpha$ for the thermal
  equilibrium of the anharmonic oscillator (\ref{eq:anharmosc}). The
  top panel shows the predictions of the classical approximation,
  while the bottom panel is the result of numerically evaluating the
  partition function. $\gamma=1/40$, and $T=2,1/2,1/5,1/15$, and
  $1/50$ for the solid, dashed, dotted, dot-dashed, and long-dashed
  curves, respectively. The upper family of lines in the top panel are
  at $ 4(\gamma\beta)=(\alpha\beta)^2$, indicating the boundary of the
  classical fluctuation-dominated regime. The lower family are
  $\alpha=T$, indicating the range of validity of the classical
  approximation in the normal regime. The circles on the lower panel
  are the (discrete) values taken by $g^{(2)}(0)$ at $T=0$. }
\label{fig:quantumgtwo}
\end{figure}

In Fig\ \ref{fig:quantumgtwo} we plot $g^{(2)}(0)$ calculated
numerically for (\ref{eq:anharmosc}) as a function of $\alpha$, at
several temperatures. The tuning parameter $\alpha$ is proportional to
density (at fixed temperature) for the polariton problem, and
generically is the pump rate in the laser model.  The mean-field
ordered phase is to the right, where we recover asymptotically
$g^{(2)}(0) \to 1$ as expected for a classical macroscopic field.  In
the top panel the classical (high-temperature) result is shown, with
different curves corresponding to different temperatures. As
temperature is lowered the result approaches more and more closely
the mean-field theory, which is of course a step function.  For
specificity we have chosen $\gamma = 1/N = 1/40$ for this
demonstration, though the qualitative evolution will be the same for
different values of $\gamma$.

While at high temperatures($\gg \gamma$) $g^{(2)}(0)$ resembles the
classical form, it is very different at low temperatures, $\lesssim
\gamma$. Well below the transition the dominant contribution to
$g^{(2)}(0)$ describes the thermal excitations of one and two photon
states, so \begin{equation} g^{(2)}(0) \approx 2e^{-2\beta\gamma}.
\end{equation} Although in this regime the statistics are strongly
sub-Poissonian, the intensity is also very low, $\approx e^{-\beta
  \alpha}$.  On crossing the transition the intensity increases and
$g^{(2)}(0)$ approaches $1$. At the lowest temperature $g^{(2)}(0)$ is
a small constant below the transition, which then approximately
follows the zero-temperature result above the transition:
$g^{(2)}(0)=1-1/n$ with $n$ the nearest integer to
$-\alpha/(2\gamma)-1/2$. At higher temperatures the quantum
corrections decrease, and the form of $g^{(2)}(0)$ is intermediate
between the classical result and the low-temperature one.

\begin{figure}[ht]
\begin{center}
\includegraphics[width=3in]{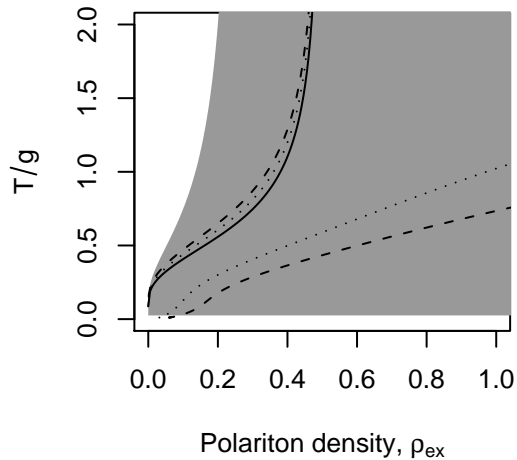}\end{center}
\caption{Fluctuation-dominated regimes in terms of density and
  temperature for the polariton condensation transition in the model
  (\ref{eq:dickeham}), with $E(i)=\omega$ so that all excitons are
  resonant with the cavity mode. The solid line is the mean-field
  boundary, and dashed(dotted) lines mark the boundaries of the
  fluctuation-dominated regions for $N=10$(40).  Curves correspond to
  those in Fig.\ \protect\ref{fig:critregionsmu}. The shading
  indicates the regions in which the static approximation holds for
  $N=10$. }
\label{fig:critregions}
\end{figure}

While the results are straightforwardly exposed in terms of the
anharmonic oscillator, the mapping of the polariton condensate to the
parameters $\alpha$ and $\gamma$ is more complex.  The essential
details are exposed in the phase diagram of Figure
\ref{fig:critregions}.

The solid line is the mean-field phase boundary between the
uncondensed (to the left) and condensed phases. Here the polariton
density is measured per site, which explains why the mean-field
boundary asymptotes to $\rho_{\mathrm{ex}} = 1/2$ (the boundary of
inversion) at large temperatures.  The dashed and dotted lines mark
regimes on either side of the mean-field transition where we estimate
fluctuations to be substantial. These lines correspond to the upper
panel of lines in Figure\ \ref{fig:quantumgtwo}.  The strong asymmetry
of the fluctuation regime about the mean-field line arises because we
have chosen to expose the results with density as a parameter, rather
than an external tuning parameter (here, naturally, the chemical
potential).  On the condensed side of the transition the number
fluctuations are large because the chemical potential for polaritons
becomes nearly clamped.  The shading marks the estimated regime of
validity of the classical approximation (akin to the lower panel of
lines in Figure\ \ref{fig:quantumgtwo}). Over most of the figure the
shaded regime encloses the fluctuation regime, and the classical
approximation (which guarantees that $g^{(2)}(0) > 1$) holds.  For
small densities and temperatures, however, we see that there is a
strongly-interacting quantum regime.

\section{Model}
\label{sec:model}

Mean-field theories of polariton condensation have now been developed
from models of a range of systems, including those with propagating
photons\cite{keeling05,keeling04}. Here we consider the simplest model
which leads to a theory of polariton condensation, the generalized
Dicke\cite{dicke54} model
\begin{equation}\label{eq:dickeham} H=\omega \psi^\dagger \psi +
\sum_i \frac{E(i)}{2} S^{z}_i + \frac{g}{\sqrt{N}}\sum_i (S^{+}_i \psi
+ \psi^\dagger S^{-}_i). \end{equation} This is the basis of standard
laser theory, as well as of our earliest mean-field
theory\cite{easthamthesis,eastham01,eastham00} of polariton
condensation. It directly describes a system of localized excitons,
for example in quantum dots, embedded in a three-dimensional
microcavity. There is a single cavity mode, with annihilation operator
$\psi$ and energy $\omega$, dipole coupled to a set of $N$ quantum
dots or atoms of the gain medium. The state $S_i^z=+(-)1$ corresponds
to the presence (absence) of an exciton on site $i$, or to an atom in the
upper (lower) of the states coupled by the lasing transition.

We have included a factor of $1/\sqrt{N}$ explicitly in the
light-matter coupling in (\ref{eq:dickeham}), so that for a cavity
mode coupled to $N$ dots $g$ is related to the observed Rabi splitting
and not to the single-dot Rabi splitting. This scaling is formally
necessary because we will be concerned with the properties of the
model (\ref{eq:dickeham}) for large $N$, so need the model to be well
behaved in the limit $N\to\infty$. But we stress that with our
convention changing the number of dots in a cavity of fixed volume
corresponds to changing both $N$ and $g$.

To construct a theory of Bose condensation of polaritons from
(\ref{eq:dickeham}) one studies its thermodynamics, fixing the total
number of excitons and photons \begin{equation}
N_{\mathrm{ex}}=\psi^\dagger\psi+\frac{1}{2}\sum_i
\left(S_i^z+1\right). \label{eq:nexdefn}\end{equation} Thus we
consider the free energy, which can be written as the functional
integral
\begin{equation}\label{eq:gpf} \frac{F}{kT}=f=\ln \int \mathcal{D}\psi\mathcal{D}\bar{\psi}
e^{-NS_{\mathrm{eff}}},
\end{equation} where \begin{equation}  S_{\mathrm{eff}}=\int_0^\beta d\tau
\bar{\psi}(\partial_{\tau}+\tilde{\omega}_{c})\psi - \ln \det P. \label{eq:seff} \end{equation} 
Here $\psi$ is related to the real electromagnetic field amplitude
$\psi_0$ by $\psi_0=\psi\sqrt{N}$, while $\ln \det P$ corresponds to
the free energy of the quantum dots in the field $\psi(\tau)$.
  
Eq. (\ref{eq:gpf}) is the free energy in the grand-canonical ensemble.
The constraint on $N_{\mathrm{ex}}$ is dealt with on average, by
introducing a chemical potential $\mu$ which appears in
(\ref{eq:seff}) as a shift of the photon and exciton energies: $\omega
\to \tilde\omega=(\omega-\mu)$ and $E(i) \to \tilde E(i)=(E(i)-\mu)=2
\tilde\varepsilon_i$.  In the limit $N\to\infty$ the relative
fluctuations of $N_{\mathrm{ex}}$ which occur in the grand-canonical
ensemble are negligible, and the grand-canonical and canonical
ensembles are expected to be equivalent. However, we are now
considering fluctuations in a finite system, for which the predictions
of the canonical and grand-canonical ensembles could differ. The
correct ensemble then depends on how the polariton number is
constrained experimentally. We consider an idealized limiting case in
which this is done by coupling to a reservoir, so that the
grand-canonical ensemble is appropriate.

We can develop an asymptotic expansion of the free energy for large
$N$ by expanding $S_{\mathrm{eff}}$ around its static saddle points
$\psi(\tau)=\psi$. The saddle-point equation gives the mean-field
estimate of the phase boundary between the normal state ($\psi=0$) and
the condensed state ($\psi$ finite). Evaluating (\ref{eq:gpf}) on the
stable saddle point leads to the mean-field estimate of the free
energy; this is the only extensive contribution to the free energy, so
that the mean-field theory gives the exact asymptotic form for the
free energy in the thermodynamic limit $N\to\infty$.

Polariton condensation in the model (\ref{eq:dickeham}) can be viewed
as a generalization of the ferroelectric transition discovered by Hepp
and Lieb\cite{hepp73,eastham01}. There has been recent interest in the
physics associated with that ferroelectric transition in a finite
model at zero temperature\cite{emary03,vidal05}. This differs from the
present problem in that there is no constraint on the polariton
number, \textit{i.e.}  $\mu=0$, and as a consequence the rotating-wave
approximation implicit in (\ref{eq:dickeham}) cannot be made. This
leads to qualitative differences\cite{emary03} in the dynamics of the
unconstrained models at $T=0$.

\section{General form of the Free energy near the transition}

\label{sec:freeenergy}

Away from the mean-field phase transition the saddle-point expansion
provides a systematic approximation scheme for the free energy of the
finite system. However, this scheme fails in the vicinity of the
mean-field transition due to the diverging occupations of the soft
fluctuations. In this section, we shall calculate the general form of
the free energy of a large but finite system near the mean-field
transition. To simplify the notation we shall take all the excitons to
have the same energy, $E(i)=E_g$, but the resulting forms may be
straightforwardly generalized to allow for a distribution of exciton
energies.

We consider the action obtained by expanding (\ref{eq:seff}) to fourth
order in $\psi$, \begin{eqnarray} S_{\mathrm{eff}}&=&S_0+\beta
\sum_{\omega}\bar{\psi}(\omega)\psi(\omega) \frac{(i\omega +
E_+)(i\omega + E_-)}{(i\omega+2\tilde{\varepsilon})}
\nonumber \\ && +
\frac{g^4}{2} \sum_{\omega_1+\omega_2=\omega_3+\omega_4}
\bar\psi_{\omega_1}\bar\psi_{\omega_2}\psi_{\omega_3}\psi_{\omega_4}
V_{\omega_1\omega_2\omega_3\omega_4} \nonumber \\ && \quad +
\ldots. \label{eq:quarticaction}\end{eqnarray} $S_0$ is the action of the two-level
systems in the absence of photons, while the remaining part describes
photons in the medium of the two-level systems. There are resonances
at $E_{\pm}$, which are the polariton energies relative to the
chemical potential
\begin{equation}\label{eq:polenergies} E_\pm = \frac{1}{2} (\tilde\omega_c +
2\tilde\varepsilon \pm
\sqrt{(\tilde\omega_c-2\tilde\varepsilon)^2+4g^2 \tanh \left(\beta
\tilde\varepsilon\right)}), \end{equation} while $V_{1234}$ is the
photon-photon interaction mediated by the two-level systems, \begin{multline} V_{1234}=\sum_{\omega_f} \{
  (i\omega_f-\tilde\varepsilon)^{-1}[i(\omega_f+\omega_3)+\tilde\varepsilon]^{-1} \\
  \times [i(\omega_f+\omega_3-\omega_2)-\tilde\varepsilon]^{-1} \\ \times
  [i(\omega_f+\omega_4+\omega_3-\omega_2)+\tilde\varepsilon]^{-1}\}.\label{eq:intsum}\end{multline}
  $\omega_{1}$, etc. are bosonic Matsubara frequencies, while
  $\omega_{f}=(n+3/4)2\pi T$ is a fermionic frequency, shifted to
  take account of the two-level
  constraint\cite{popov88,keeling05}.

\subsection{Static free energy}

As we approach the transition from the normal side one of the
polariton energies $E_{\pm}$ is approaching zero, and perturbation
theory fails. Over most of this fluctuation regime, however,
$\omega_1=2\pi T$ is large compared with the energy of the soft mode.
The dominant contribution to the free energy for a large system close
to the transition then comes from the static (classical) paths.
Retaining only these dominant contributions gives an approximation for
(\ref{eq:gpf}) near the transition,
\begin{eqnarray}
\label{eq:staticfullcfenergy}f&\approx& \ln \int d\psi d\psi^*
e^{-N(\beta
\tilde\omega_c|\psi|^2-\ln\cosh\beta\sqrt{\tilde{\varepsilon}^2+g^2|\psi|^2})}
\\ &=& f_0+\ln \int d\psi d\psi^\ast e^{-N(a |\psi|^2+b |\psi|^4 + c
|\psi|^6 + \ldots)}. \label{eq:staticfenergy}\end{eqnarray} Here
$a=\beta E_+E_-/(2\tilde\varepsilon)$ is the static part of the Gaussian
kernel in (\ref{eq:quarticaction}), measuring the distance to the
transition, $b$ is the static part of the interaction term, and $c, d,
\ldots$ are higher-order interaction strengths which do not depend on the
system size $N$.  $f_0=N \ln \cosh \beta\tilde\varepsilon$ is the free
energy of the two-level systems in zero field. The explicit form for
$b$, obtained either from (\ref{eq:intsum}) or by expanding the
exponent in (\ref{eq:staticfullcfenergy}), is
\begin{equation}b=g^4 \beta \frac{\tanh(\beta
\tilde\varepsilon)-\beta\tilde\varepsilon\sech^2(\beta\tilde\varepsilon)}{8\tilde\varepsilon^3}. \label{eq:staticintstrength}
\end{equation}

\subsection{Reduction to an oscillator}

\label{sec:freeenergy-osc}

At low temperatures the non-perturbative regime may lie outside the
regime of validity of the static approximation, so that time-dependent
paths must be considered. The divergences of perturbation theory are
still at small frequencies, so for a large system close to the
transition we can replace the action with its low frequency form.
Considering for definiteness the region near the transition where
$E_-$ vanishes, we have $\omega, E_-$ as small parameters, while
generically $E_+$ and $\tilde\varepsilon$ are finite. The Gaussian
term in (\ref{eq:quarticaction}) can then be straightforwardly
approximated as
\begin{equation}  \frac{\beta E_+}{2\tilde\varepsilon}
  \sum_{\omega}\bar{\psi}(\omega)\psi(\omega)
  (i\omega+E_-), \label{eq:gausslowfreq}\end{equation} with corrections proportional to
  the ratios of small to finite parameters. Approximating the
  interaction term is more involved, because the result of the
  summation (\ref{eq:intsum}) takes different forms depending on how many of
the external frequencies coincide. This problem can be avoided by
  restricting our attention to the low-temperature regime $T \ll
  \tilde \varepsilon$, in which all the forms lead to the same
  low-frequency approximation
\begin{equation} \frac{g^4 \beta}{8\tilde\varepsilon^3}
  \sum_{\omega_1+\omega_2=\omega_3+\omega_4}
  \bar\psi_1\bar\psi_2\psi_3\psi_4.
  \label{eq:quartlowfreq}\end{equation} The corrections to
(\ref{eq:quartlowfreq}) are again small in terms of the ratios of
small to large parameters, \textit{e.g.} $\omega/\tilde\varepsilon$ and
$T/\tilde\varepsilon$. Since the coefficient in
(\ref{eq:quartlowfreq}) is the low-temperature limit of $b$, it is
convenient to replace it with $b$, leading to a low-energy
approximation to the original theory which is valid at the quantum
level for low temperatures, and at the classical level at higher
temperatures.

(\ref{eq:gausslowfreq}) and (\ref{eq:quartlowfreq}) become, after
rescaling the fields $\psi\to\psi\sqrt{\frac{2\tilde\varepsilon}{E_+
    N}}$, the action for a quantum anharmonic oscillator
(\ref{eq:anharmosc}).  The oscillator frequency $\alpha=E_-$, and
the interaction strength is
$\frac{b}{\beta N}\left(\frac{2\tilde\varepsilon}{E_+}\right)^2$, or
$\frac{g^4}{2\tilde\varepsilon N E_+^2}$ at low
temperatures. 

\subsection{Validity of the static approximation}

\label{sec:freeenergy-stat}

The static approximation leading to (\ref{eq:staticfullcfenergy})
holds because the energy of fluctuations is much less than
temperature. On the normal side of the transition this only occurs in
a region close to the transition. However, the behavior of the
excitation spectrum on the condensed side is different. This
spectrum\cite{eastham01,eastham00,easthamthesis} comprises a mode
which is always at zero frequency relative to $\mu$, and two modes
which at the transition are at positive and negative of the
non-vanishing polariton energy. Thus there appears to be no mode which
becomes soft at the transition, and so the classical approximation
does not appear to be controlled by the distance from the transition.

This puzzle is resolved by inspecting the anharmonic oscillator
(\ref{eq:anharmz}), which we introduced earlier in section
\ref{sec:outline}.  On the normal side of the transition the
nonlinear terms in (\ref{eq:anharmz}) are irrelevant, and the
parameter controlling the classical approximation is $\beta \alpha$.
On the condensed side the exponent in (\ref{eq:anharmz}) can be
rewritten in terms of the number fluctuations $\delta n=n -
n_{min}=n-(-\alpha/(2\gamma))$ as $-\beta \gamma (\delta n)^2$. Thus
the parameter controlling the static approximation on the condensed
side is $\beta \gamma$. Returning to the polariton model close to the
transition where $E_-$ vanishes, we see that the classical
approximation holds for the low-energy fluctuations when
\begin{equation}
  \frac{b}{N}\left(\frac{2\tilde\varepsilon}{E_+}\right)^2 \ll 1,
  \label{eq:cclassin} \end{equation} 
provided the temperature is small compared with the Rabi splitting and
$\tilde\varepsilon$. At higher temperatures we expect an inequality of
the same general form, $TN/g\gtrsim 1$, but with numerical
differences due to the renormalization of the effective interaction by
the occupation of the high-energy polariton and the frequency
dependence of the interaction.

Our previous computations\cite{eastham01,eastham00,easthamthesis} of
the fluctuation spectrum were done at a Gaussian level,
\textit{i.e.} approximating the partition function of the fluctuations with
that of a harmonic oscillator. This approximation predicts that the
characteristic frequency of fluctuations about the condensate is zero
because there is no linear term in $\delta n$ when the exponent of
(\ref{eq:anharmz}) is expanded around a finite $n$. To obtain a finite
level spacing one must go beyond the Gaussian approximation and
include interactions, which generate a finite level spacing $\sim
1/N$. This may be contrasted with the normal state, where the harmonic
oscillator part of (\ref{eq:anharmz}) is enough to generate a finite
level spacing $\alpha$.

\section{Correlation functions and phase diagram}

\label{sec:corrfuns}

The correlation functions of the cavity field can be obtained by
differentiating (\ref{eq:staticfullcfenergy}) with respect to
$\tilde\omega_c$ or (\ref{eq:staticfenergy}) with respect to $a$. Note
that the field is classical: the integrand of
(\ref{eq:staticfullcfenergy}) can be interpreted as a probability
distribution for the intensity of the cavity field. Non-classical
fields are associated with time-dependent paths, which give complex
integrands in the path integral (\ref{eq:gpf}) that cannot be
interpreted as classical probabilities.

The integral (\ref{eq:staticfullcfenergy}) is only tractable
numerically. However, the interactions $c, d, \ldots$ do not affect
the asymptotic behavior of the correlation functions as $N\to\infty$
if $a\geq 0$ and $b>0$, because the quartic nonlinearity restricts
fluctuations of the field to $|\psi|\lesssim
N^{-1/4}b^{-1/4}$. Therefore we can obtain the asymptotic forms of the
correlation functions on the normal side of the mean-field phase
boundary from (\ref{eq:staticfenergy}) with $c, d, \ldots=0$. The
corresponding free energy is, discarding additive constants,
\begin{eqnarray} f&=&\ln \int d\psi d\psi^\ast
  e^{-N(a|\psi|^2+b|\psi|^4)}\label{eq:fquadintgl} \\ &=&\ln
\frac{e^{\frac{Na^2}{4b}}\erfc\left(\frac{a}{2}\sqrt{\frac{N}{b}}\right)}{\sqrt{bN}}.
\label{eq:fquadexplicit} \end{eqnarray}%% \frac{Na^2}{4b}+\ln \erfc \left(\frac{a}{2}\sqrt{\frac{N}{b}}\right)
%% \end{equation}

On the condensed side where $a<0$ we may still use the approximation
(\ref{eq:fquadintgl}) so long as the order parameter $|\psi|$ is
small. It is a weaker approximation than on the normal side, because
the minimum of the exponent in (\ref{eq:staticfenergy}) occurs for
$|\psi|^2\sim 1$, so that truncation to a quartic theory produces
errors in the leading asymptotics of the correlation
functions. However, these errors are proportional to the order
parameter, and so are numerically small close enough to the
transition, even if they are not asymptotically small in $N$.

We now present explicit results for the behavior of the cavity field
in the model (\ref{eq:gpf}) when $N$ is finite but large. For
orientation, the mean-field phase boundaries $E_{\pm}=0$ are shown as
the solid lines in Fig.\ \ref{fig:critregions}.  $\rho_{ex}$ is the
number of polaritons per site, $\langle N_{\mathrm{ex}}/N \rangle$
with $N_{\mathrm{ex}}$ given by (\ref{eq:nexdefn}); note that this
differs from the definition of $\rho_{ex}$ used in Refs.\ 
\onlinecite{eastham00,eastham01,easthamthesis} by a shift of $0.5$.
Temperature is expressed in units of $g$, which is one-half of the
collective Rabi splitting at resonance(see Eq.\ \ref{eq:polenergies}).

For completeness we begin by considering the region far from the
transition, where the correlation functions can be obtained using the
saddle-point expansion. On the normal side of the mean-field
transition there is no saddle-point contribution to the photon
density $\langle \psi^\dagger \psi \rangle$, so the first
non-vanishing contribution is at order $1/N$. This term is just the
expectation value of the photon density from the Gaussian part of
(\ref{eq:quarticaction}), \textit{i.e.} the photon density in a population of
non-interacting polaritons with energies $E_{\pm}$:
\begin{eqnarray}
\label{eq:photongaussian} \langle \psi^\dagger \psi \rangle &=& \frac{(E_+ -
2\tilde{\varepsilon})n_{\mathrm{B}}(E_+) + (2\tilde{\varepsilon} -
E_-)n_{\mathrm{B}}(E_-)}{N(E_+-E_-)}, \\ n_\mathrm{B}(x)& =&
\frac{1}{e^{\beta x}-1}.\end{eqnarray} Since at this order in $N$ we
have non-interacting particles the many-photon correlation functions
are related to the photon number by Wick's theorem. In particular, for
the static part of the correlation function measured by Deng \textit{et al.} we
have
$g^{(2)}(0)=\langle\psi^\dagger\psi^\dagger\psi\psi\rangle/\langle\psi^\dagger\psi\rangle^2=2+O(1/N)$. 

On the condensed side of the mean-field transition the leading-order
contributions to the correlation functions come from the saddle
point. Thus $\langle \psi^\dagger \psi \rangle \sim 1$, and we expect
$g^{(2)}(0)=1+O(1/N)$. Calculations of the higher-order terms are
complicated due to the presence of the zero mode, and so we shall
not pursue them here.

In the region close to the mean-field transition the saddle-point
expansion must break down, to allow the correlation functions to
smoothly interpolate between their forms in the two states. We can
estimate the boundaries of this crossover region by equating the
magnitudes of successive terms in the large-N expansion of the photon
density, $\langle \psi^\dagger \psi \rangle$. For the approximate form
(\ref{eq:fquadexplicit}) we see that the crossover regime obeys
$Na^2\lesssim 4 b$, or
\begin{equation}\label{eq:critinequality}\frac{2g^4}{\beta N
\tilde\varepsilon
E_+^2E_-^2}\left(\tanh(\beta\tilde\varepsilon)-\beta\tilde\varepsilon\sech^2(\beta\tilde\varepsilon)\right)
\gtrsim 1.\end{equation} Note that the numerical factor in this
expression is not meaningful, since it depends on the precise
definition of the boundaries of the crossover regime.

\begin{figure}[ht]
\begin{center}
\includegraphics[width=3in]{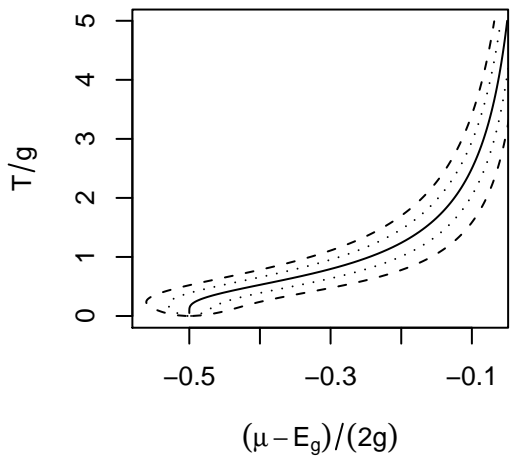}\end{center}
\caption{Fluctuation-dominated regimes for the polariton condensation
  transition with $\Delta=0$. Solid lines are the mean-field
  boundaries.  Dashed (dotted) lines mark the boundaries of the
  fluctuation-dominated regions for $N=10$(40).}
\label{fig:critregionsmu}
\end{figure}

In Fig.\ \ref{fig:critregionsmu} we show the phase diagram of the
system as a function of $\mu$ and $T$, for $\Delta
g=\omega-E_g=0$. The solid lines are the phase boundary of the
infinite system, $E_+E_-=0$, while the remaining lines are the
boundaries of the crossover regions (\ref{eq:critinequality}) for
systems of 10 and 40 quantum dots. One general feature which can be
seen in this figure is that the size of the fluctuation regions scales
as $1/\sqrt{N}$, as can be deduced from (\ref{eq:fquadexplicit}). The
particular shapes of the fluctuation regions regions come from the
interplay between the thermal occupations and the strength of the
interaction, which is temperature and density dependent. In
particular, the present model has the unusual feature that the
fluctuation regions vanish in the high-temperature limit, where the
limits of the inequality (\ref{eq:critinequality}) are approximately
\begin{equation}
  4\tilde\varepsilon=g^2\beta-2\Delta g \pm
  \frac{2\beta g^2}{\sqrt{3N}}. \end{equation} Physically, this
§occurs because  for large $T$ the
number of fluctuations at fixed chemical potential behaves as
$n_{\mathrm{fluc}}=n_{\mathrm{B}}(E_+)+n_{\mathrm{B}}(E_-) \sim
T$ while the interaction strength $b$ vanishes like
$1/T^4$. Thus the mean interaction energy $\langle n_{\mathrm{fluc}}^2
b \rangle$ vanishes in the high temperature limit. The occupation of
the fluctuations diverges only as $T$ because they are confined
to a finite energy range, so that increasing temperature does not
increase the number of relevant fluctuation modes, as it would in the more
familiar case of a semi-infinite band of states.  The interaction
vanishes because at high temperatures the field does not affect the
occupation of the two-level systems: the free energy of a two-level
system, which appears in the exponent of
(\ref{eq:staticfullcfenergy}), becomes independent of field $|\psi|$
as $T\to \infty$. 

Fig.\ \ref{fig:critregions} shows the same phase diagram in terms of
density and temperature, obtained by relating chemical potential to
density using the mean-field results. For the normal state we use
\begin{equation}
\rho_{\mathrm{ex}}=\rho_{\mathrm{excitons}}=(1-\tanh(\beta\tilde\varepsilon))/2,
\label{eq:mfdenseq}\end{equation} while for the condensate the quartic
theory gives \begin{equation}
  \rho_{\mathrm{ex}}=\rho_{\mathrm{excitons}}+kT\frac{\partial}{\partial
    \mu}\left(\frac{a^2}{4b}\right).\label{eq:mfdenseqcond}
\end{equation} These relations introduce qualitative differences
between the fluctuation-dominated regions in this figure and those in
Fig.\ \ref{fig:critregionsmu}. In particular, the
fluctuation-dominated region becomes much larger on the condensed
side, because the chemical potential is only weakly dependent on
density in the condensate.

The shading on Fig.\ \ref{fig:critregions} indicates the regimes of
validity of the classical approximation, combining the analysis of
sections\ \ref{sec:freeenergy-osc} and\ \ref{sec:freeenergy-stat} with
Eqs.\ \ref{eq:mfdenseq} and\ \ref{eq:mfdenseqcond}. On the normal side
of the mean-field transition we shade the region $\beta E_-<1$. On the
condensed side the analogous inequality is (\ref{eq:cclassin}).
However, the numerical factors in this result are only correct close
to the phase boundary. Since this only occurs at low densities we
have plotted the inequality $T/g>1/(4N)$, corresponding to the
low-density limit of (\ref{eq:cclassin}).

\begin{figure}[ht]
\begin{center}
\includegraphics[width=3in]{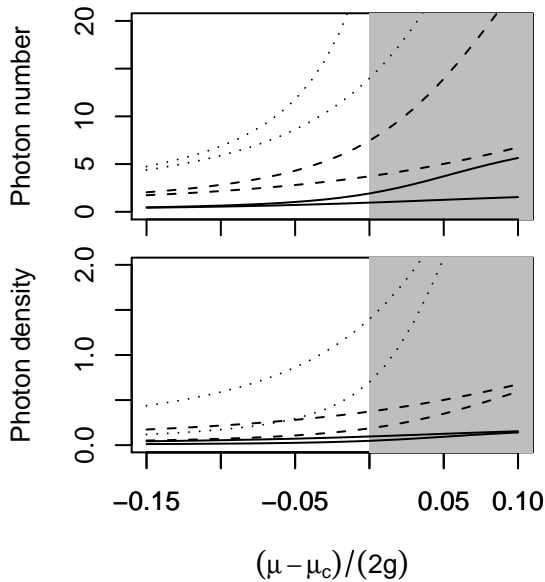}
\end{center}
\caption{Photon number $N\langle \psi^\dagger \psi \rangle$ (top
panel) and density $\langle \psi^\dagger \psi \rangle$ (bottom panel)
as functions of $(\mu-\mu_c)/(2g)$ for $\Delta=0$. $T/g=0.25$ (solid
lines), 0.75 (dashed lines), and 1.5 (dotted lines). $N=10$ (lower line
of each pair in the top panel, upper in the bottom panel) and $N=40$
(upper line in top panel, lower in bottom panel). Shading marks the
condensed region of the mean-field theory.}
\label{fig:photofmu}
\end{figure}

In the upper panel of Fig.\ \ref{fig:photofmu} we plot the prediction
of (\ref{eq:fquadexplicit}) for the number of photons in the the
cavity
\begin{equation} N \langle \psi^\dagger\psi\rangle =
-\frac{\partial f}{\partial a},
\end{equation} as a function of the deviation of $\mu$ from its
mean-field critical value $\mu_c$. We take $\Delta=0$, and plot curves
for $T/g=0.25, 0.75$ and $1.5$, and for $N=10$ and $40$. 

In general each of the $N$ two-level systems makes a contribution of
order $1/N$ to the number of cavity photons, due to the scaling of the
coupling constant. Well below the transition we could neglect the
interactions between the photons and approximate the integrand of
(\ref{eq:staticfenergy}) by a Gaussian. This gives $N \langle
\psi^\dagger \psi \rangle\approx 1/a$. In the Gaussian regime each
two-level system contributes incoherently to the cavity field, so the
total photon number is of order $1$; such scaling is demonstrated by
the collapse of the curves corresponding to different system sizes in
the left side of the top panel of Fig.\ \ref{fig:photofmu}.

As we increase the chemical potential through the mean-field
transition the occupancy of the cavity field increases, and the
interactions begin to generate coherence between the contributions of
the different two-level systems. Far in the condensed state this
coherence is complete: all the $\sim N$ two-level systems contribute
coherently to the cavity field, giving a photon number which scales as
$N$. This scaling can be seen on the right of the lower panel of Fig.\
\ref{fig:photofmu}, which shows the photon number per two-level
system. The order of $N$ contribution to the photon number comes from
the saddle point of (\ref{eq:fquadintgl}), and is $-a/(2b)$. This
saddle-point contribution is the exponential in
(\ref{eq:fquadexplicit}); it survives in the condensed state, but is
canceled by the asymptotic expansion of the error function well into
the normal state.

In the region near the transition neither the Gaussian nor
saddle-point approximations are appropriate, and the full form
(\ref{eq:fquadexplicit}) must be used. There is partial coherence
amongst the two-level systems, leading to a photon number which scales
as $\sqrt{N}$. Explicitly, we find
$N\langle\psi^\dagger\psi\rangle=\sqrt{N/(\pi b)}$ for the photon
number at the transition. The scaling can be shown from a more general
argument by noting that the $a$ dependent part of
(\ref{eq:fquadintgl}) is a function of $\sqrt{N}a$ and $b$, so at the
transition
\begin{eqnarray} N\langle \psi^\dagger \psi \rangle = -\left.\frac{\partial
    f(\sqrt{N}a,b)}{\partial a}\right|_{a=0} \nonumber \\ =
    -\sqrt{N}g(0,b). \end{eqnarray}

\begin{figure}[ht]
\begin{center}
\includegraphics[width=3in]{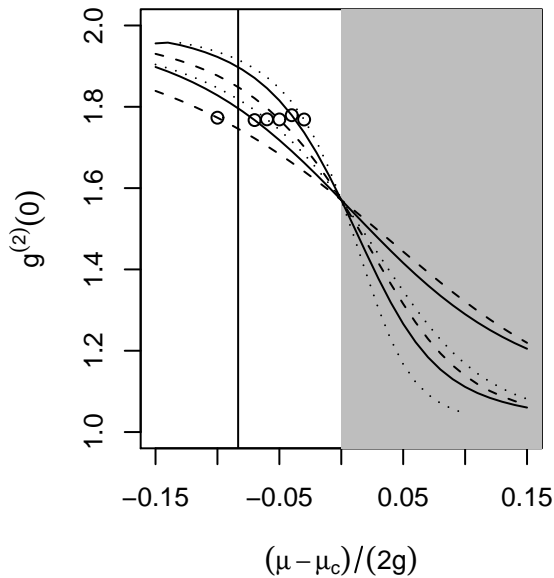}
\end{center}
\caption{Intensity-intensity correlation function $g^{(2)}(0)$ of
  the cavity photons as a function of chemical potential
  $(\mu-\mu_c)/(2g)$ for $\Delta=0$. $T/g=0.25$ (solid lines),
  0.75 (dashed lines), and 1.5 (dotted lines).  $N=10$ (lower line of
  each pair in the normal region) and $N=40$ (upper line). Shading
  marks the condensed region of the mean-field theory.  Circles
  correspond to the lines marking the fluctuation-dominated regimes of
  the normal state in Fig.\ \protect\ref{fig:critregions}. The vertical line
  corresponds to the limit of the shading in the normal state on Fig.\ 
  \protect\ref{fig:critregions} for $T/g=0.25$; the analogous lines for
  $T/g=0.75$ and $1.5$ are off the scale.}
\label{fig:iicorrfun}
\end{figure}

Fig.\ \ref{fig:iicorrfun} illustrates the results obtained from
(\ref{eq:fquadexplicit}) for the static intensity-intensity
correlation function
\begin{equation} g^{(2)}(\tau=0)=1+\frac{\frac{\partial^2f}{\partial
a^2}}{\left(\frac{\partial f}{\partial a}\right)^2}. \end{equation}
Well below the transition $g^{(2)}(0)$ approaches the value of $2$
associated with a thermal mixture of non-interacting photons, while
well above it $g^{(2)}(0)$ approaches $1$, corresponding to the bunched
photons of \textit{e.g.} a coherent state or large-amplitude number state. The
crossover occurs over a range of chemical potentials which scales as $1/\sqrt{N}$. 

\section{Laser fluctuations}

\label{sec:laser}

In this section, we briefly review the standard theory of fluctuations
in a single-mode laser close to threshold, following the work of
Risken\cite{risken70}. We shall see that the photon statistics
predicted by this theory are similar to those predicted by the theory
of polariton condensation. This similarity holds in spite of the fact
that the theory of condensation describes a system in thermal
equilibrium while laser theory describes one which is not in thermal
equilibrium.

The canonical laser theories are based on the dynamics of models such
as (\ref{eq:dickeham}) in the presence of dissipation processes. Such
dissipation processes can be formally generated by coupling each dot,
and the photon mode, to an oscillator bath. Thus one considers
Hamiltonians of the general form
\begin{equation} H_{T}=H+H_{\mathrm{res}}+H_{\mathrm{ph-res}}+H_{\mathrm{s-res}}, \label{eq:laserham}
\end{equation} where $H=H_{\mathrm{ph}}+H_{\mathrm{s}}+H_{\mathrm{s-ph}}$ is given by Eq. (\ref{eq:dickeham}),
$H_{\mathrm{res}}$ describes the harmonic oscillator baths, and the remaining
two terms couple the system and baths. The standard model is to couple
the cavity mode to its bath with Hamiltonian
\begin{equation} H_{\mathrm{ph-res}}= \sum_p g_p(\psi^\dagger d_p + d^\dagger_p
\psi), \end{equation} and the dots to their baths with Hamiltonian
\begin{equation} H_{\mathrm{s-res}}= \sum_i \sum_p g^{\bot}_{pi}(d_{pi}^\dagger
S_i^-+S_i^+ d_{pi}) + g^{\|}_{ip}d_{pi}^\dagger d_{pi}
S_z. \end{equation} 

To develop dynamical equations for the system variables alone one takes
the Liouville equation for the evolution of the density matrix and
averages over the reservoir variables. The interaction between the
system and reservoir variables is treated using second-order
perturbation theory, leading to an integro-differential equation for
the time evolution of the reduced density matrix $\rho$. This equation
is then approximated by a first-order differential equation,
implicitly neglecting memory effects, formally by having Markovian
baths whose spectra are completely white. 

Finally, one integrates the atomic degrees of freedom out of the
density matrix for the system, assuming that the atomic system is very
strongly damped. This leads to an equation of motion for the $P$
representation of the field density matrix, $W(\psi,\psi^\ast,t)$.
Neglecting terms associated with the quantum nature of the
field\cite{risken70} this equation becomes the Fokker-Planck equation
describing diffusion in a quartic potential:
\begin{eqnarray}\frac{\partial W}{\partial t}+\zeta
\frac{\partial}{\partial\psi} (d-|\psi|^2)\psi W + \zeta
\frac{\partial}{\partial\psi^\ast} (d-|\psi|^2)\psi^\ast W \nonumber
\\ = 4q\frac{\partial^2 W}{\partial\psi\partial\psi^\ast}.
\label{eq:laserfpunscaled}
\end{eqnarray}
The coefficients in this form are the strength of the nonlinearity
$\zeta$(denoted $\beta$ in Ref.\ \onlinecite{risken70}), the linear
gain or loss $\zeta d$, and a diffusion constant $q$. 

To compare with the theory of fluctuations near polariton
condensation we make the same rescaling of the field as we did there,
$\psi \to \psi\sqrt{N}$. In terms of this rescaled field the
steady-state probability distribution obtained from
(\ref{eq:laserfpunscaled}) is proportional to the integrand of
(\ref{eq:fquadintgl}). The parameters are related to the laser
parameters as $a=-\zeta d/(2q)$ and $b=\zeta N/(4q)$. The scaling of
these terms follows from the microscopic expressions for $\zeta, q$
and $d$ given in Eq. (2.17, 2.27) of Ref.\ \onlinecite{risken70}.
Substituting $g\to g/\sqrt{N}$ so that $g$ is as defined in Eq.\ 
\ref{eq:dickeham}, and noting that the total inversion $\sigma$ scales
as $N$, we find $q \sim N^0$, $\zeta \sim N^{-1}$, $\zeta d \sim N^0$.
Therefore, as in the polariton condensate, the parameters $a$ and $b$
are independent of $N$.

Thus in the classical approximation the photon statistics of a laser
and a condensate differ only due to the dependence of $a$ and $b$ on
the microscopic parameters of the system. This is similar to the usual
universality which occurs close to a second-order phase transition,
where the singular parts of the observables are given by universal
scaling functions, with non-universal relations between the scaling
parameters and physical parameters such as temperature. In the finite
systems considered here the observables are not singular, but for
large $N$ the correlation functions in the two problems are identical
functions of a parameter describing the distance from the transition.
This arises because the collective behavior of the electromagnetic
field in both theories is described by a quartic ``free-energy'' , and
the system size $N$ enters these free energies in the same way.

Within our approximations, then, the only scope for making a
distinction between polariton condensation and lasing arises from the
forms of $a$ and $b$ in the two systems. These forms are
model-dependent, and expected to change in more realistic models of
either system. However, we note that there are differences in the
physics of these forms in the laser and the condensate, which could be
expected to be robust.  For example, in the laser the noise strength
$q$ is the spontaneous emission rate into the lasing mode, while in
the condensate the analogous noise strength is temperature.  It
appears that the laser amplifies spontaneous emission noise, while the
condensate amplifies thermal noise. The parameters $a$ and $b$ also
contain different physics: in the laser $a$ is associated with gain or
loss and $b$ with gain depletion, while in the condensate $a$ is a
single-particle energy and $b$ an interaction strength. Thus the size
of the threshold regime in absolute units is controlled by different
physics in the two theories.

The analogy between the laser threshold and a second-order phase
transition is well-known, and previous authors have pointed out that
this analogy extends beyond the thermodynamic
limit\cite{rice94,haken75}: in a finite system the singularities
associated with a phase transition are rounded in the same way that
the sharp singularities of the lasing transition in a rate-equation
treatment are rounded by noise. We see that for the laser model
(\ref{eq:laserham}) the parameter controlling this rounding is $N$,
the system size, \textit{exactly} as in the usual thermodynamic case.  This
appears to differ from the picture obtained from classical stochastic
models\cite{rice94}, in which the corresponding parameter is the
fraction of the spontaneous emission directed into the lasing mode.

\section{Quantitative estimates for current experiments}
\label{sec:numbers}

We can use our model to obtain some indication of the scale of the
fluctuation-dominated region in current experimental systems. Such
systems are generally planar microcavities, rather than the
three-dimensional cavity of (\ref{eq:dickeham}). However the
luminescence which may be evidence for condensation is localized,
suggesting that the condensate itself is localized, presumably due to
some combination of kinetics, self-trapping, and disorder. In a
condensate localized on scale $L$ we expect there to be an energy cost
$\sim (L^2 m_p^2)^{-1}$ associated with spatial variations, where $m_p
\sim 10^{-5} m_e$ is the in-plane mass of the polaritons. If this
energy cost is large compared with temperature we expect spatial
fluctuations to be unimportant, and the present theory to apply.

Let us consider in particular the data discussed by Weihs, Deng, Snoke,
and Yamamoto in Ref.\ \onlinecite{weihs2004}. These experiments are done on
a GaAs planar microcavity with twelve quantum wells and a Rabi
splitting of $14.9 \mathrm{meV}$, so $g\approx 86\mathrm{K}$. Their
condensate is typically $5 {\mathrm{\mu m}}$ in diameter, giving a
temperature scale for spatial fluctuations of $60\mathrm{K}$. We take
the Mott density, $10^{12} \mathrm{cm^{-2}}$, to be an upper bound on
the density of available exciton sites, so the condensate of diameter
$5 {\mathrm{\mu m}}$ corresponds to $N\sim 10^5$.

The temperature in these systems is extremely low relative to the Rabi
splitting: $T=5K$ corresponds to $T/g=0.06$. At these low temperatures
the critical density is not associated with the single-mode phase
transition studied here but with spatial fluctuations, whose effect on
the mean-field theory is analyzed in Refs.\ \onlinecite{keeling04} and
\onlinecite{keeling05}. Thus our theory cannot describe the transition
that is being crossed in the experiments. However, once the condensate
has formed we expect it to apply, since the condensate which forms is
apparently localized, and the spatial degrees of freedom are frozen
out. We estimate that the achieved condensate densities of
$10^{10}\mathrm{cm^{-2}}$ correspond to $\rho_{\mathrm{ex}} \approx
10^{10}/{n_{\mathrm{Mott}}}=10^{-2}$. For this density the transition
temperature of our theory is $T/g=0.2 \approx 20\mathrm{K}$. For
$N=10^{5}$ the fluctuation-dominated region extends over around
$20\mathrm{mK}$ on the normal side of the transition and $1\mathrm{K}$
on the condensed side. The experimental temperature is far outside
this region, so the present theory suggests that the fluctuations of
the condensate are negligible and the light from the condensate area
is almost perfectly coherent.  If raising the temperature did not
change the localization then this would allow the sharp crossover
predicted here to be observed. This complication would be absent for
systems with external in-plane confinement of the polaritons, such as
pillar microcavities\cite{vahala03}.

For a system in thermal equilibrium, we have argued that the quantum
regime occurs when the temperature is smaller than the characteristic
level spacing produced by the photon-photon interaction. For the
particular model used here this is when $TN \sim g$. We estimate that
this gives a scale of $1\mathrm{mK}$ for the systems of Ref.\ 
\onlinecite{weihs2004}, which is far below the achieved temperatures.
Furthermore, these are open systems coupled to baths. In such a
system, the effect of dissipation is loosely to broaden the level
spectrum through the decay process.  To have quantum effects that
dominate for the longest times one then requires linewidths that are
narrow in comparison to the level spacing. This is not achieved, since
a temperature scale of $1\mathrm{mK}$ corresponds to a linewidth on the order
of $10 \mu \mathrm{eV}$, far below the linewidth associated with the decay of
the cavity mode.

\section{Discussion}
\label{sec:discussion}

Recently, three groups\cite{yoshie04,reithmaier04,peter05} have achieved
strong-coupling of single quantum dots in microcavities, and are thus
beginning to approach the regime, already achieved in atom cavity
optics\cite{thompson92}, of a single two-level system strongly coupled
to a cavity mode. Clearly the nonlinear regime\cite{mckeever03} of
such a system is dominated by quantum effects. However, we note that
our model suggests that this single-atom limit is not the only way to
see strong quantum effects. For a cavity of fixed volume one should
replace $g/\sqrt{N}$ with $g$ in (\ref{eq:dickeham}). The effective
photon-photon interaction then scales as $N$(whereas it scales as
$N^{-1}$ at fixed dot density), so that quantum effects survive to
higher temperatures when the number of dots increases in a cavity of
fixed volume: a situation opposite to the normal thermodynamic limit
we consider elsewhere, in which quantum effects move to lower
temperatures as the system size increases. In this context it is
interesting to note the experiments of Ref.\ \onlinecite{rempe91}, in
which the quantum statistical properties of light emitted from a
weakly driven atom-cavity system are measured. The observed
$g^{(2)}(0)$ was below the classical limit of 1, and furthermore its
value was apparently independent of the number of atoms in the cavity.
This would naturally arise if the noise increased as $N$, as it does
in standard laser theory, canceling the increase in the level
spacing.

The behavior of a finite system in which spatial fluctuations are
allowed could be analyzed using renormalization group
arguments\cite{cardybook}. We expect it to depend on the interplay
between system size and an effective co-ordination number or
interaction range: a small system with a large
co-ordination (long-range interactions) will be dominated by
finite-size fluctuations over a larger region of the phase diagram
than it is dominated by spatial fluctuations, whereas the reverse will
occur for a large system with a small co-ordination (short-range
interactions). In the former case we can view the finite-size
fluctuations as corrections to mean-field theory, as here. This cannot
be true in the latter case, however, since the theory of the finite
system should then involve critical exponents different from those of
mean-field theory.

The phase diagram of the infinite two-dimensional system, allowing for
spatial fluctuations, has been studied in Ref.\ 
\onlinecite{keeling05}. The deviations between the phase boundary
there and that of mean-field theory indicate the regions in which
spatial fluctuations of an infinite system are significant. Since the
photon mass is very small, effectively providing a long-range
interaction between excitons, these deviations are only significant at
very low densities, where even a long-range interaction gives a small
co-ordination number. Hence, except at low densities, we expect even a
relatively large system to have mean-field like finite-size behavior.

\section{Conclusions}
\label{sec:conclusions}

In this paper we have analyzed the behavior of a model of a finite
polariton condensate close to the mean-field transition, and shown
that it is formally similar to the theory describing a laser close to
the laser threshold. This similarity fails when the polariton
condensate reaches a low-temperature quantum regime, since the laser
theory is classical -- notwithstanding that the noise in the laser
theory might ultimately have its origins in quantum effects.

In the classical regime we find that the intensity distribution for
the photons in a finite polariton condensate is of the same form as
that obtained from conventional laser theory. Thus the photon
statistics are not expected to reveal any fundamental difference
between a condensate and a laser, but of course the parametrization is
different and the scales have different meanings. In particular, for a
condensate in thermal equilibrium the sharpness of the crossover is
controlled by temperature, whereas in a laser it is controlled by the
noise introduced by coupling to external baths.

Quantum effects dominate in general when the level spacing exceeds the
noise strength. We have seen that for the polariton condensate in
equilibrium the relevant level spacing is the photon-photon
interaction, and the relevant noise strength is temperature. In the
conventional thermodynamic limit of our model the temperature scale
for quantum effects decreases with system size, but the opposite
occurs when increasing the dot number with fixed cavity volume. Hence
it is not always the smallest systems which have the most strongly
quantum-mechanical collective behavior.

\begin{acknowledgments}
  
  We thank Simon Kos, Vikram Tripathi, Jonathan Keeling, and Mike Gunn
  for helpful discussions of this work, and acknowledge support from
  the EPSRC, Sidney Sussex College, Cambridge, and the EU RTN Project
  No.  HPRN-2002-00298.

\end{acknowledgments}

%\bibliography{paper}

\end{document}